\documentclass[9pt,twocolumn,twoside]{opticajnl}
\journal{opticajournal} 

\setboolean{shortarticle}{false}

\usepackage{lineno}
\usepackage{textcomp}

\usepackage{layouts}
\usepackage{mathptmx}
\usepackage{etoolbox}
\usepackage{comment}
\usepackage{amsmath}
\usepackage{textcomp}
\usepackage{hyperref}

\title{Backward-Wave Difference-Frequency Generation in Thin-Film Lithium Niobate}

\author[1, $\dagger$]{Yesim Koyaz}
\author[1, 2, $\dagger$]{Ozan Yakar}
\author[3, $\dagger$]{Furkan Ayhan}
\author[4]{Victor Brasch}
\author[3]{Luis Guillermo Villanueva}
\author[1, *]{Camille-Sophie Brès}

\affil[1]{Ecole Polytechnique Fédérale de Lausanne, Photonic Systems Laboratory (PHOSL), 1015 Lausanne, Switzerland}
\affil[2]{Current address: Deutsches Elektronen-Synchrotron DESY, Notkestr. 85, 22607 Hamburg, Germany}
\affil[3]{Ecole Polytechnique Fédérale de Lausanne, Advanced Nano-electromechanical Systems Laboratory (ANEMS), 1015 Lausanne, Switzerland}
\affil[4]{Q.ANT GmbH, Handwerkstraße 29, 70565 Stuttgart, Germany}
\affil[*]{e-mail: camille.bres@epfl.ch}
\affil[$\dagger$]{These authors contributed equally.}

\doi{ } 

\begin{abstract}

Second-order nonlinear processes involving counter-propagating light generation have been explored for quantum applications and optical parametric oscillators. However, realizing these processes on integrated photonic platforms such as thin-film lithium niobate~(TFLN) remains challenging because of the extremely short quasi-phase matching~(QPM) periods required, which are difficult to achieve with standard fabrication workflows. In this work, we achieve 1425~nm periodicity on 800~nm-thick X-cut TFLN via poling prior to etching. We present the first integrated demonstration of backward-wave difference-frequency generation~(BWDFG), using a pump near 775~nm and a counter-propagating signal near 1980~nm, and thereby combine the high nonlinear efficiency of QPM with favorable dispersion characteristics. Within the same waveguide, we demonstrate backward-wave second-harmonic generation~(BWSHG), where the pump and second-harmonic are counter-propagating, together with BWDFG. Idler generation spans from 1244~nm to 1290~nm and simulations predict extended coverage up~to $\sim$2200 nm. This process provides broad spectral tunability while remaining tolerant of fabrication-induced dimensional offsets, addressing a key challenge for the TFLN platform.

\end{abstract}

\setboolean{displaycopyright}{false} 

\begin{document}

\maketitle

\thispagestyle{plain}
\pagestyle{plain}

\section{Introduction} \label{sec:intro}
Conventional second-order frequency conversion involves the pump, signal, and idler waves all propagating in the same forward direction through the nonlinear medium. In contrast, counter-propagating and backward configurations, require one or more of the generated waves to travel in the opposite direction to the pump, fundamentally altering the phase-matching conditions. Such configurations have been investigated for quantum applications such as spontaneous parametric down-conversion~(SPDC), since they enable spatial separation of the generated signal and idler photons \cite{lancoSemiconductorWaveguideSource2006, luoCounterpropagatingPhotonPair2020, liuObservationFrequencyuncorrelatedPhoton2021, kellnerCounterpropagatingSpontaneousParametric2026}, and for mirrorless optical parametric oscillators because of their intrinsic feedback mechanism \cite{harrisPROPOSEDBACKWARDWAVE1966, dingBackwardOpticalParametric1996, canaliasMirrorlessOpticalParametric2007, laurellBackwardwaveOpticalParametric2025}. These applications require sufficiently high optical power to either surpass the detector's noise floor or exceed an intrinsic threshold, a constraint that integrated platforms can relax through the enhanced field confinement of nanophotonic waveguides. However, realizing the quasi-phase-matching~(QPM) condition required for these processes is particularly challenging, owing to their significantly larger phase mismatch compared to their conventional forward-propagating counterparts. Within integrated platforms, backward second-harmonic generation~(BSHG), where the second-harmonic propagates counter to the pump, was first demonstrated in the $\rm Si_{3}N_{4}$ platform via optically induced periodic $\chi^{(2)}$, enabling first-order QPM of the process \cite{yakarIntegratedBackwardSecondHarmonic2023}. Yet, the value of optically induced effective $\chi^{(2)}$ is orders of magnitude lower than the intrinsic $\chi^{(2)}$ of ferroelectric materials, limiting the overall conversion efficiency.

\begin{figure*}[!ht]
\centering
\includegraphics[width=184.2mm]{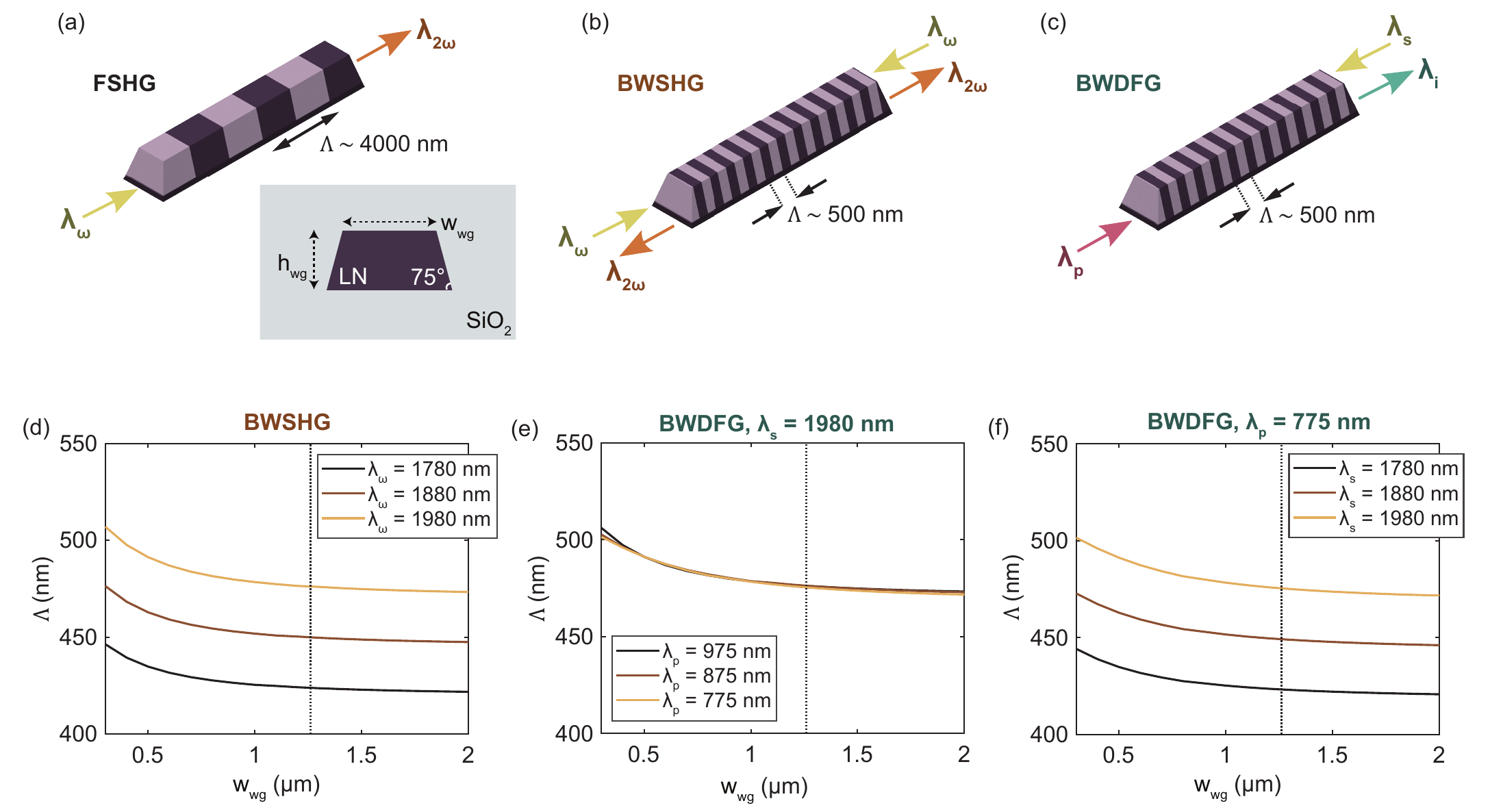}%
\caption{\label{fig1}(a~-~c)~Schematic of the investigated nonlinear processes, namely (a)~forward second-harmonic generation~(FSHG), (b)~backward-wave second-harmonic generation~(BWSHG), (c)~backward-wave difference-frequency generation~(BWDFG). Inset in (a): Cross-section of the investigated waveguide. (d~-~f)~Calculated QPM periods~($\Lambda$) as a function of $w_{\rm wg}$ (d)~for BWSHG at different pump wavelengths ($\lambda_{\omega}$) and (e)~for BWDFG at different pump wavelengths~($\lambda_{\rm p}$) and (f)~for BWDFG at different signal wavelengths~($\lambda_{\rm s}$). The dashed line corresponds to the arbitrarily selected cross-section $w_{\rm wg}$~=~1.26~\textmu m, which is further investigated in Fig.~\ref{fig2}.}%
\end{figure*}

Thin-film lithium niobate~(TFLN) is a platform that combines high intrinsic $\chi^{(2)}$~($\sim$ 50 pm/V \cite{shojiAbsoluteScaleSecondorder1997}) with low-loss nanophotonic waveguides with tight confinement \cite{wuLongLowLossLitiumNiobate2018, zhuTwentynineMillionIntrinsic2024}. It facilitates periodic poling through high-voltage ms pulses, with QPM can be achieved via lithographically defined domain inversion. In recent years, conventional forward-propagating second-harmonic generation~(FSHG) in TFLN has been extensively studied, requiring poling periods of around $\sim$4000~nm, which are sensitive to waveguide dimensions and operating wavelength~(Fig.~\ref{fig1}(a)) \cite{wangUltrahighefficiencyWavelengthConversion2018, jankowskiUltrabroadbandNonlinearOptics2020, zhuIntegratedPhotonicsThinfilm2021, heftiFabricationtolerantModalPhase2025, chenAdaptedPolingBreak2024}. In contrast, backward-wave second-harmonic generation~(BWSHG), where counter-propagating pump inputs generate a second-harmonic~(SH) signal, requires poling periods on the order of 500~nm~(Fig.~\ref{fig1}(b)), imposing significantly tighter requirements on the poling process \cite{sabattiNanodomainPolingUnlocking2025, yangSymmetricSecondharmonicGeneration2024, yakarBackwardWaveSecondHarmonic2025}. In Z-cut TFLN, such short periods are more easily achieved through vertical domain inversion across the film thickness \cite{yangSymmetricSecondharmonicGeneration2024}, whereas in X-cut TFLN, they are harder to realize because of the competing lateral domain growth as the poling depth increases, thus requiring more precise poling pulse engineering \cite{younesiPeriodicPolingMicrometerrange2021, nagySubmicrometerPeriodicPoling2020, zhaoPolingThinfilmXcut2020} or alternative poling techniques such as sidewall poling \cite{frankenMilliwattlevelUVGeneration2026, sabattiNanodomainPolingUnlocking2025} or electron-beam poling \cite{fergestadHighResolutionElectronBeamPoling2025}. Using these approaches, BSHG and BWSHG have been demonstrated in the C-band using X-cut TFLN \cite{sabattiNanodomainPolingUnlocking2025, yakarBackwardWaveSecondHarmonic2025, ferfergestadCounterpropagatingNondegenerateFrequency2025}.

In this work, we demonstrate BWSHG and then extend the analysis to backward-wave difference-frequency generation~(BWDFG), where pump and signal inputs counter-propagate~(Fig.~\ref{fig1}(c)), to investigate the spectral tunability of counter-propagating processes. Broadband difference-frequency generation~(DFG) using conventional forward DFG~(FDFG) has experimentally achieved a spectral coverage of up to $\sim$300~nm in 600~nm-thick TFLN-on-silica \cite{koyazUltrabroadbandTunableDifference2024} and $\sim$1000~nm on sapphire \cite{mishraUltrabroadbandMidinfraredGeneration2022}, indicating films with thicknesses of $\rm \geq$600~nm to be more suitable for broadband operation. This is further supported by OPA demonstrations achieving $\sim$600~nm of bandwidth with an input near 2~\textmu m \cite{ledezmaIntenseOpticalParametric2022} in comparably thick TFLN. However, the QPM wavelength in FDFG is highly sensitive to deviations in waveguide dimensions, requiring extensive scans of signal wavelength~($\lambda_{\rm s}$) and pump wavelength~($\lambda_{\rm p}$) to identify the phase-matched idler wavelength~($\lambda_{\rm i}$) after fabrication. Our analysis shows that BWDFG is intrinsically more tolerant against such variations while still allowing broadband tunability.

Our analysis of BWSHG and BWDFG is carried out using periodically poled, fully-etched, $800$~nm-thick X-cut TFLN waveguides. We achieve poling periods as short as $1425$~nm, enabling third-order QPM for these processes. We present a numerical analysis of the poled waveguides, complemented by experimental validation, including an assessment of their robustness against variations in waveguide dimensions. We realize both processes within the same periodically poled waveguide and explore the limits of spectral tunability by operating far from the degenerate regime (i.e.~$\lambda_{\rm s}~=~2\lambda_{\rm p}$), by placing $\lambda_{\rm s}$ and $\lambda_{\rm i}$ near the Thulium~band~($\sim$1980~nm) and O~band~($\sim$1270~nm), respectively. In addition, separate demonstrations of $\lambda_{\rm p}$~tuning (from 765~nm to 781~nm) and temperature tuning of the generated idler are presented, enabling $\lambda_{\rm i}$ to span from 1244~nm to 1290~nm, limited by the available sources.

\section{Waveguide design \& dispersion engineering} \label{sec:sim} 

\begin{figure*}[!ht]
\centering
\includegraphics[width=184.2mm]{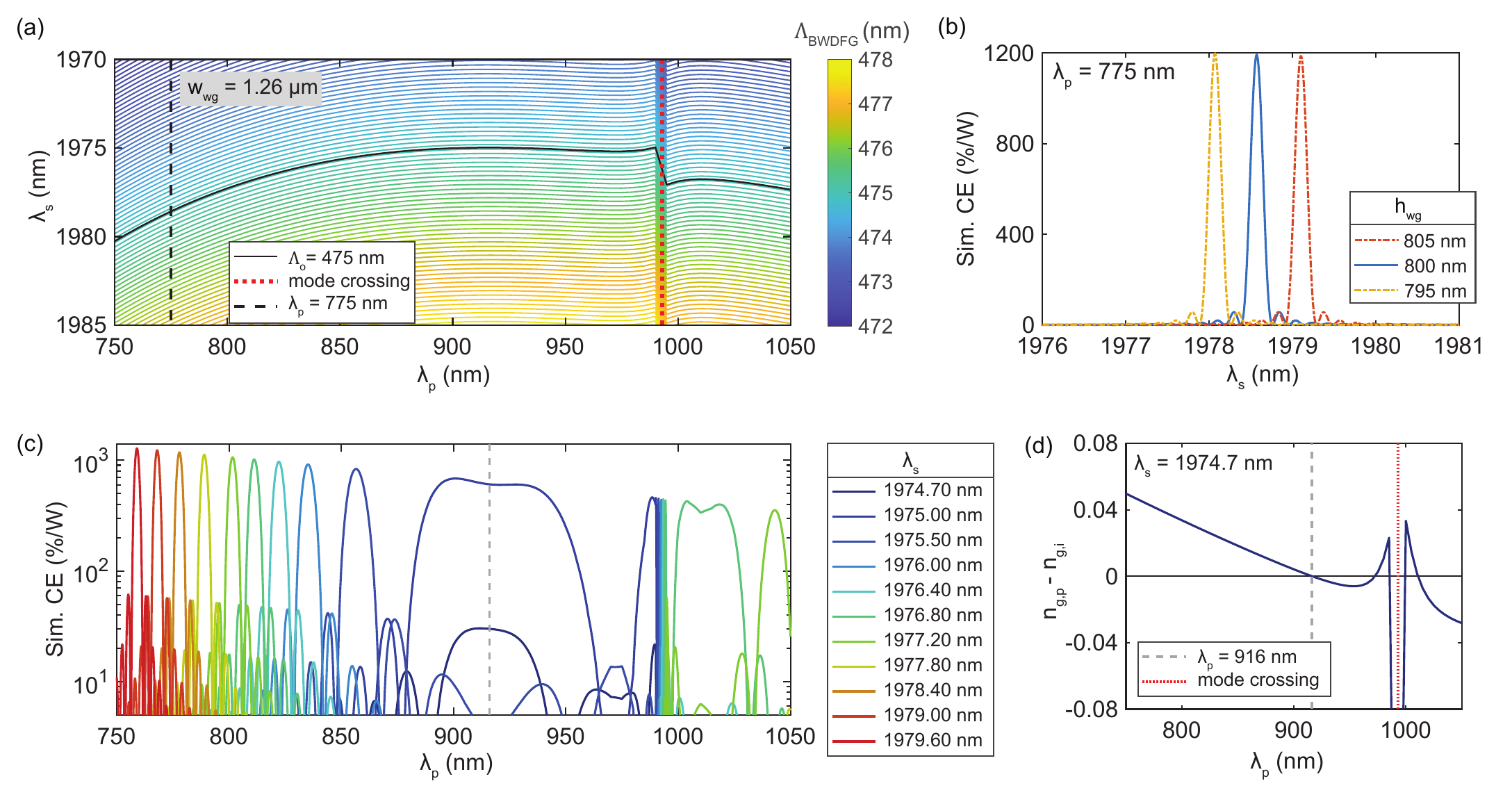}%
\caption{\label{fig2} (a)~Calculated backward-wave difference-frequency generation (BWDFG) poling period~($\Lambda_{\rm BWDFG}$) as a function of ($\lambda_{\rm s}$) and ($\lambda_{\rm p}$) for $w_{\rm wg}$~=~1.26~\textmu m. The mode-crossing wavelength~($\rm \sim 993$~nm) and $\lambda_{\rm p}$ corresponding to the measurement band~($\rm \sim 775$~nm) are shown by the red dotted line and black dashed lines respectively. The contour corresponding to the reference QPM period~($\Lambda_{\rm o}$~=~475~nm) is shown as a solid black line. (b)~Simulated BWDFG conversion efficiency~(CE) versus $\lambda_{\rm s}$ at constant $\Lambda_{\rm o}$ for a target waveguide height of $h_{\rm wg}$~=~800~nm and possible deviations, namely 795~nm and 805~nm. (c)~Simulated BWDFG~CE versus $\lambda_{\rm p}$ at the same $\Lambda_{\rm o}$ for several $\lambda_{\rm s}$. (d) Group index mismatch between pump and idler for $\lambda_{\rm s}$~=~1974.7~nm.}%
\end{figure*}

Our numerical analysis focuses on the QPM requirements of BWSHG and BWDFG for fundamental TE mode interactions for the cross-section shown in Fig.~\ref{fig1}(a)-inset with a fixed $h_{\rm wg}$ of 800~nm. The first-order QPM period~($\Lambda$) is calculated using \eqref{eq:period}:

\begin{equation}
\label{eq:period}
{\Lambda_{\rm BWSHG}~=~\frac{\lambda_{\omega}}{2n_{\mathrm{2}\omega}} \text{ and } 
\Lambda_{\rm BWDFG}~=~1 \bigg/\left\lvert\frac{n_{\rm p}}{\lambda_{\rm p}} - \frac{n_{\rm i}}{\lambda_{\rm i}} + \frac{n_{\rm s}}{\lambda_{\rm s}}\right\rvert } 
\end{equation}

where $n_{(\omega,\mathrm{2}\omega \mathrm{,p,s,i})}$ denote the effective indices and $\lambda_{(\omega, \mathrm{2}\omega \mathrm{,p,s,i})}$ are the corresponding wavelengths of the BWSHG pump and SH fields and the BWDFG pump, signal, and idler fields, respectively.

Fig.~\ref{fig1}(d) presents the dependence of $\Lambda_{\rm BWSHG}$ on waveguide width ($w_{\rm wg}$) for different $\lambda_{\omega}$ values. The results indicate that the phase-matched wavelength is relatively insensitive to variations in $w_{\rm wg}$, yet $\Lambda_{\rm BWSHG}$ shifts significantly with respect to $\lambda_{\omega}$. 

Fig.~\ref{fig1}(e) demonstrates the dependence of $\Lambda_{\rm BWDFG}$ at different $\lambda_{\rm p}$ values at a fixed $\lambda_{\rm s}$~=~1980 nm, while Fig.~\ref{fig1}(f) demonstrates the dependence of the same quantity at different $\lambda_{\rm s}$ values at a fixed $\lambda_{\rm p}$~=~775 nm. In Fig.~\ref{fig1}(f), the dependence of $\Lambda_{\rm BWDFG}$ on $\lambda_{\rm s}$ follows a trend comparable to that of $\Lambda_{\rm BWSHG}$ on $\lambda_{\omega}$~(Fig.~\ref{fig1}(d)), indicating strong tolerance with respect to variations in the $w_{\rm wg}$ value. In contrast, Fig.~\ref{fig1}(e) shows that $\Lambda_{\rm BWDFG}$ remains nearly constant for different values of $\lambda_{\rm p}$, at constant $w_{\rm wg}$, indicating the possibility of broad spectral tuning of $\lambda_{\rm i}$ through $\lambda_{\rm p}$. 
 
To further investigate tunability, in Fig.~\ref{fig2}(a), $\Lambda_{\rm BWDFG}$ is presented as a two-dimensional map of $\lambda_{\rm p}$ and $\lambda_{\rm s}$, for a reference $w_{\rm wg}$. It confirms the distinct trends of $\lambda_{\rm p}$ and $\lambda_{\rm s}$ over a larger $\lambda_{\rm p}$ range spanning 750~nm to 1050~nm. We observe a discontinuity near $\lambda_{\rm p} = 993$~nm, which is associated with the mode crossing between the fundamental TE mode and the $\rm TM_{ 10}$ mode of the investigated cross-section (red dotted line). We also highlight the measurement band, namely $\lambda_{\rm p} \sim 775$~nm, by the black dashed line. 

For this cross-section, we calculate BWDFG conversion efficiency~(CE) for perfect rectangular poling using \eqref{eq:ce}:
\begin{equation}
\label{eq:ce}
{{\rm CE}~=\frac{P_{\rm i}}{P_{\rm s}P_{\rm p}}~= \frac{4}{m^2\pi^2}\frac{L^2\omega_{\rm i}^2 \left\lvert \chi^{(2)}\iint\limits_{\rm WG} E_{\rm s}^*E_{\rm i}^*E_{\rm p}\,{\rm d}x{\rm d}y\right\rvert ^2} {2 \varepsilon_0 {\rm c}^3 \prod\limits_{k={\rm i, s ,p}} n_k \iint \limits_{\infty} \lvert E_k\rvert^2 \,{\rm d}x{\rm d}y} 
\mathrm{sinc}^2 \left(\Delta \beta L/2\right) 
}
\end{equation} 

where $w_{\rm i}$ is the idler's angular frequency, $m$ is the QPM order (with odd integers only, i.e., $m =1,3,5,...$ \cite{fejerQuasiphasematchedSecondHarmonic1992}), $\varepsilon_0$ is vacuum permittivity, $c$ is the speed of light and $\chi^{(2)}$ is the second-order susceptibility. $P_{( \mathrm{p,s,i})}$ denotes the power and $E_{(\mathrm{p,s,i})}$ denotes the electric field of pump, signal, and idler power, respectively, with ${\rm sinc}(x)={\rm sin}(x)/x$. Poled length ($L$) is 4.5 mm and the phase mismatch ($\Delta\beta$) is calculated by \eqref{eq:delta_beta} at a selected QPM period $\Lambda_{\rm o}$:

\begin{equation}
\label{eq:delta_beta}
{\Delta\beta = \beta_{\rm p} + \beta_{\rm s} - \beta_{\rm i} - \frac{2\pi}{\Lambda_{\rm o}}} {\rm with \ } {\Lambda_{\rm o} = \frac{1}{m}\Lambda_{\rm poling}}
\end{equation}

where $\beta_{\rm (p,s,i)}$ corresponds to the propagation constants of the pump, signal and idler, respectively, and $\Lambda_{\rm poling}$ is the poling period. $\Lambda_{\rm o}$~=~475~nm has been marked as a black contour line in Fig.~\ref{fig2}(a).

We highlight that first-order~($m=1$) and third-order~($m=3$) QPM differ in CE, where $m^{\rm th}$-order QPM reduces CE by a factor of $1/m^2$ relative to the first-order. The numerical analysis uses first-order~($m=1$) QPM, as it represents the ideal, highest-efficiency case, while experiments rely on third-order due to its threefold longer $\Lambda_{\rm poling}$ (discussed in more detail in Sec.~\ref{sec3}). Despite this difference, the analysis~(based on Eqs.~(\ref{eq:ce}) and (\ref{eq:delta_beta})) remains valid for predicting phase mismatch behavior, bandwidth, and spectral tunability, and can therefore be used to interpret the experimental third-order results.
 
In Fig.~\ref{fig2}(b), CE at $\lambda_{\rm p}$~=~775~nm shows a narrowband response with a bandwidth $\leq 0.2$~nm, confirming the sensitivity to $\lambda_{\rm s}$. To investigate the impact of thickness variations, we repeat the analysis for $\pm 5$~nm deviations around the nominal value of 800~nm.
Across this parameter range, the phase-matched $\lambda_{\rm s}$ shifts by only $\sim$1~nm, as a result of the counter-propagating signal input. Notably, this shift is significantly smaller than the chip-to-chip variations associated with FDFG and FSHG \cite{fergestadSimultaneousBroadbandNonlinear2026,koyazUltrabroadbandTunableDifference2024, koyazDesignRulesFrequency2024}

\begin{figure*}[htb]
\includegraphics[width=184.2mm]{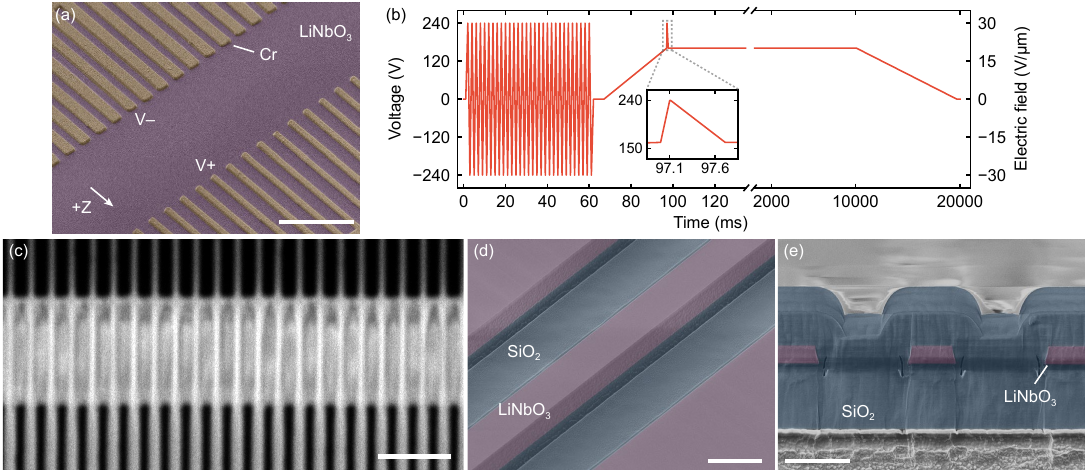}
\caption{\label{fig3} (a) False-color scanning electron microscope (SEM) image of the chromium (Cr) electrodes patterned on the lithium niobate substrate. Poling period ($\mathrm{\Lambda}$) is 1425~nm and electrode gap is 8~\textmu m. Positive (V+) electrodes are designed to have a duty cycle of 20~\% to reduce lateral domain broadening. Scale bar is 5~\textmu m. (b) High-voltage waveform applied for periodic poling, tracking both applied voltage (left axis) and corresponding electric field (right axis). The waveform consists of: (i) bipolar nucleation pulses with a period of 2~ms (ii) a rapid ramp to a peak field of 30~V/\textmu m held for 0.01~ms to initiate domain inversion, (iii) a constant hold at 20~V/\textmu m for $\sim 10$~s, and (iv) an extended 10~s fall time to prevent back-switching.
(c) Two-photon microscope (TPM) image of the resulting periodically poled domain structure. Scale bar is 5~\textmu m. (d) False-color SEM image of the fully-etched lithium niobate waveguide. Scale bar is 3~\textmu m. (e) False-color SEM image of the input facet of the lithium niobate waveguide. Scale bar is 3~\textmu m.}
\end{figure*}

Fig.~\ref{fig2}(c) investigates the $\lambda_{\rm p}$ coverage of CE for $\lambda_{\rm s}$ ranging from 1974.7~nm to 1979.6~nm. Through this $\lambda_{\rm s}$ tuning (with a range of $\leq 6$~nm), we achieve idler generation starting from $\lambda_{\rm p}$~=~750~nm~($\lambda_{\rm i}$~=~1250~nm) and reaching up to $\lambda_{\rm p}$~=~1050~nm~($\lambda_{\rm i}$~=~2200~nm). We also highlight the broadband behavior near $\lambda_{\rm p}$~=~916~nm at $\lambda_{\rm s}$~=~1974.7~nm, resulting in a full width at half maximum of more than 50~nm. To identify the underlying conditions, we calculate $\Delta \beta$~(using \eqref{eq:delta_beta}) at the phase-matched location using a first-order Taylor expansion for $\Delta\lambda_{\rm p}$ near the
phase-matched operating point with $\lambda_{\rm p,o}$ and $\lambda_{\rm i,o}$ \cite{jankowskiDispersionengineeredChi^2Nanophotonics2021, gaoWidelyTunableCavityenhanced2025}:

\begin{equation}
\label{eq:bw}
\Delta \beta(\lambda_{\rm p,o} + \Delta\lambda_{\rm p})~=~\frac{-2 \pi\Delta\lambda_{\rm p}}{\lambda_{\rm p,o}^{2}} (n_{\rm g, p} - n_{\rm g, i})
\end{equation}

where $n_{\rm g, p}$ and $n_{\rm g, i}$ correspond to the group indices of the phase-matched pump and the idler, respectively. This shows that $\Delta \beta$ remains near zero for larger $\Delta\lambda_{\rm p}$ when $n_{\rm g, p}~=~n_{\rm g, i}$, enabling broadband operation. The group index mismatch between pump and idler, plotted in Fig.~\ref{fig2}(d), confirms that this condition is satisfied near 916~nm, consistent with the bandwidth observed in Fig.~\ref{fig2}(c).

Overall, broadband spectral coverage of $\lambda_{\rm i}$ is achieved through broad tuning of $\lambda_{\rm p}$ together with fine adjustment of $\lambda_{\rm s}$, with the latter also providing resilience against variations in waveguide dimensions.

\section{Experimental Validation} \label{sec3}

To validate the proposed analysis, BWSHG measurements were performed near 1980 nm, and BWDFG measurements were carried out by coupling near-775~nm~($\lambda_{\rm p}$) and near-1980~nm~($\lambda_{\rm s}$) light into waveguides. To facilitate the QPM condition of these processes, we performed periodic poling using 4.5~mm long top-electrodes prior to waveguide etching. 

We highlight that the optimal poling depth of this TFLN thickness~($\sim 800$~nm) is over three times larger than the optimal lateral width~($\sim 250$~nm), corresponding to 50\% of the calculated first-order QPM period, $\Lambda_{\rm BWSHG/DFG}$~($\sim 500$~nm). 

To operate with a more balanced aspect ratio, we employ third-order QPM, i.e., $m=3$, with $\Lambda_{\rm poling}$~=~1425~nm~($\Lambda_{\rm o}$~=~475~nm). In this case, optimal lateral width becomes comparable to the film thickness, enabling improved poling depth through relaxed restrictions on domain shape. 

\subsection{Short-Period Poling} \label{sec:pole}

We control the size of the poled domains primarily through the poling pulse parameters and electrode geometry. We employ narrow positive electrodes~(20\% duty cycle) paired with wider negative electrodes~(50\% duty cycle), separated by a gap of $8$~\textmu m, as shown in Fig.~\ref{fig3}(a). The narrow positive electrodes facilitate confined domain nucleation, while the electrode gap enables sufficient poling depth with limited lateral growth \cite{sabattiNanodomainPolingUnlocking, rosenmanDomainBroadeningQuasiphasematched1998}. 

The applied poling waveform is shown in Fig.~\ref{fig3}(b). Bipolar triangular nucleation pulses with a period of 2~ms were first applied to promote homogeneous domain formation and minimize missing domains \cite{nagySubmicrometerPeriodicPoling2020}. Subsequently, the main poling pulse was applied, characterized by a fast rise time to a peak electric field of 30~V/\textmu m which was sustained for a duration of 0.01~ms. These parameters were chosen to suppress lateral domain growth, aiming to preserve an ideal 50\% duty cycle. To maximize poling depth despite the short peak duration, the field was held at 20~V/\textmu m for $\sim 10$~s before concluding with an extended fall time of 10~s to enhance domain stabilization \cite{sheltonRobustPolingFrequency2025}.

The resulting two-photon microscope (TPM) image~(Fig.~\ref{fig3}(c)) reveals that the periodically poled domains exhibit dark edges and a light gray center along the electrode gap, indicating partial but dominant inversion across the film thickness \cite{rusingSecondHarmonicMicroscopy2019}. On this basis, the previously described electrode configuration and waveform were used in the fabrication of the investigated waveguides, while the remainder of the fabrication was carried out as detailed in \cite{ayhanFabricationPeriodicallyPoled2025}. Fig~\ref{fig3}(d) presents the scanning electron microscope (SEM) image of the waveguide, while Fig.~\ref{fig3}(e) shows the waveguide's input facet after silica cladding deposition and substrate release.

\begin{figure}[!ht]
\includegraphics[width=92.1mm]{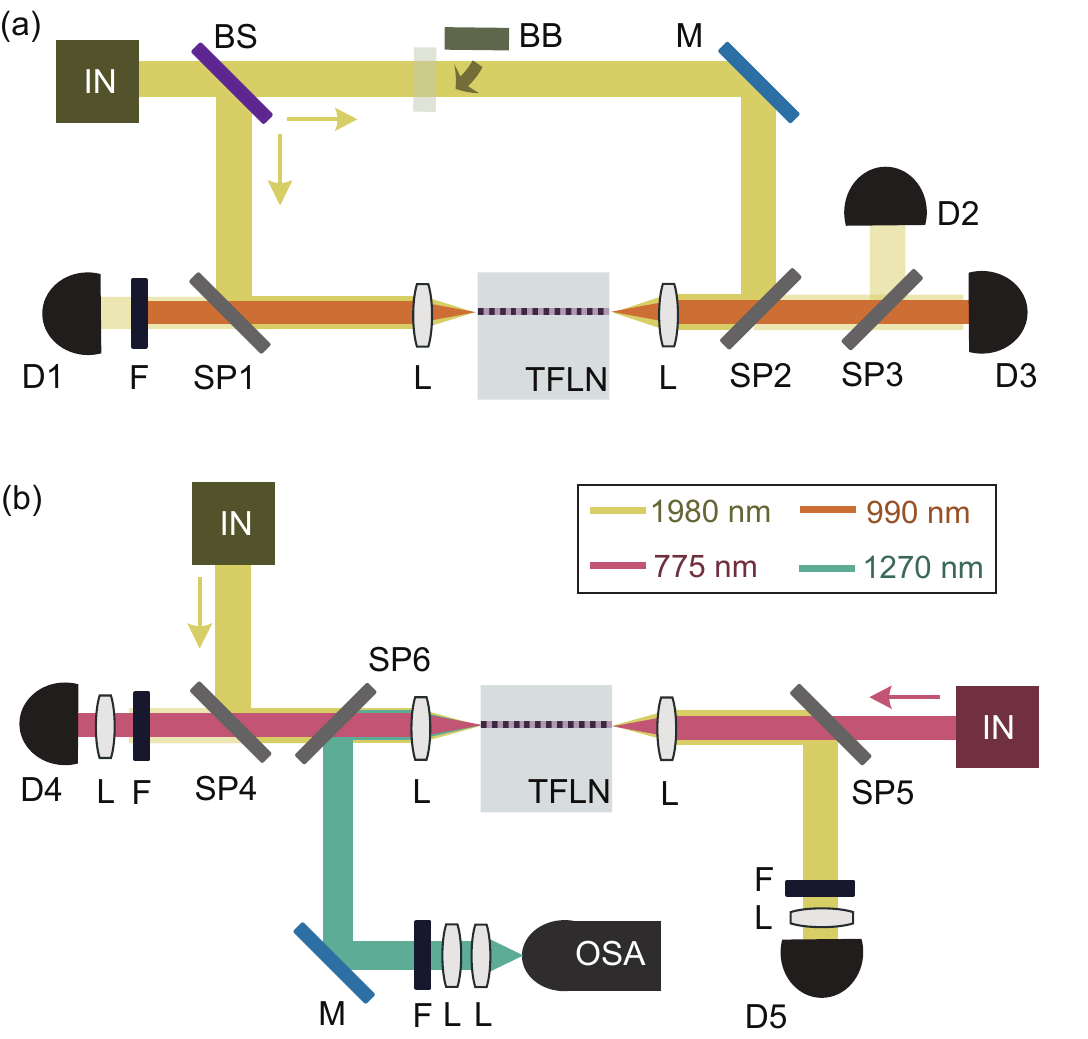}
\caption{\label{fig4} Optical characterization setup for (a) backward-wave second-harmonic generation (BWSHG) operating with $\lambda_{\omega}$ near 1980~nm and $\lambda_{2\omega}$ near 990~nm and (b) backward-wave difference-frequency generation (BWDFG) with inputs near 1980~nm and 775~nm and the generated idler near 1270~nm. BS:~beam splitter, BB:~beam Block, M:~mirror, L:~lens, F:~filter, SP:~short-pass dichroic mirror, OSA:~optical spectrum analyzer, IN:~frequency-doubled quasi-CW laser at 775~nm, a tunable CW laser from 765~nm to 781~nm, or an amplified tunable CW laser from 1900~nm to 2000~nm.}
\end{figure}

\subsection{Characterization Setup} \label{sec:setup} 

The setup used for BWSHG characterization is shown in Fig.~\ref{fig4}(a) where the chip is placed on a temperature-controlled stage. A~continuous wave laser~(Sacher LION), operating near $1980$~nm, is connected to a thulium-doped fiber amplifier. The collimated beam is then split into two by a beam splitter~(BS) and the resulting beams are coupled into the waveguide from both ends after reflecting through short-pass dichroic mirrors~(SP1 and SP2). An additional SP~(SP3) is added to the right arm to separate the residual pump from the generated SH~(near $990$~nm). The outcoupled pump on both arms and the generated SH are recorded simultaneously through individual power meters~(D1, D2 and D3). To characterize FSHG, a beam block~(BB) is inserted in the right arm after the beam splitter.

\begin{figure}[!b]
\includegraphics[width=92.1mm]{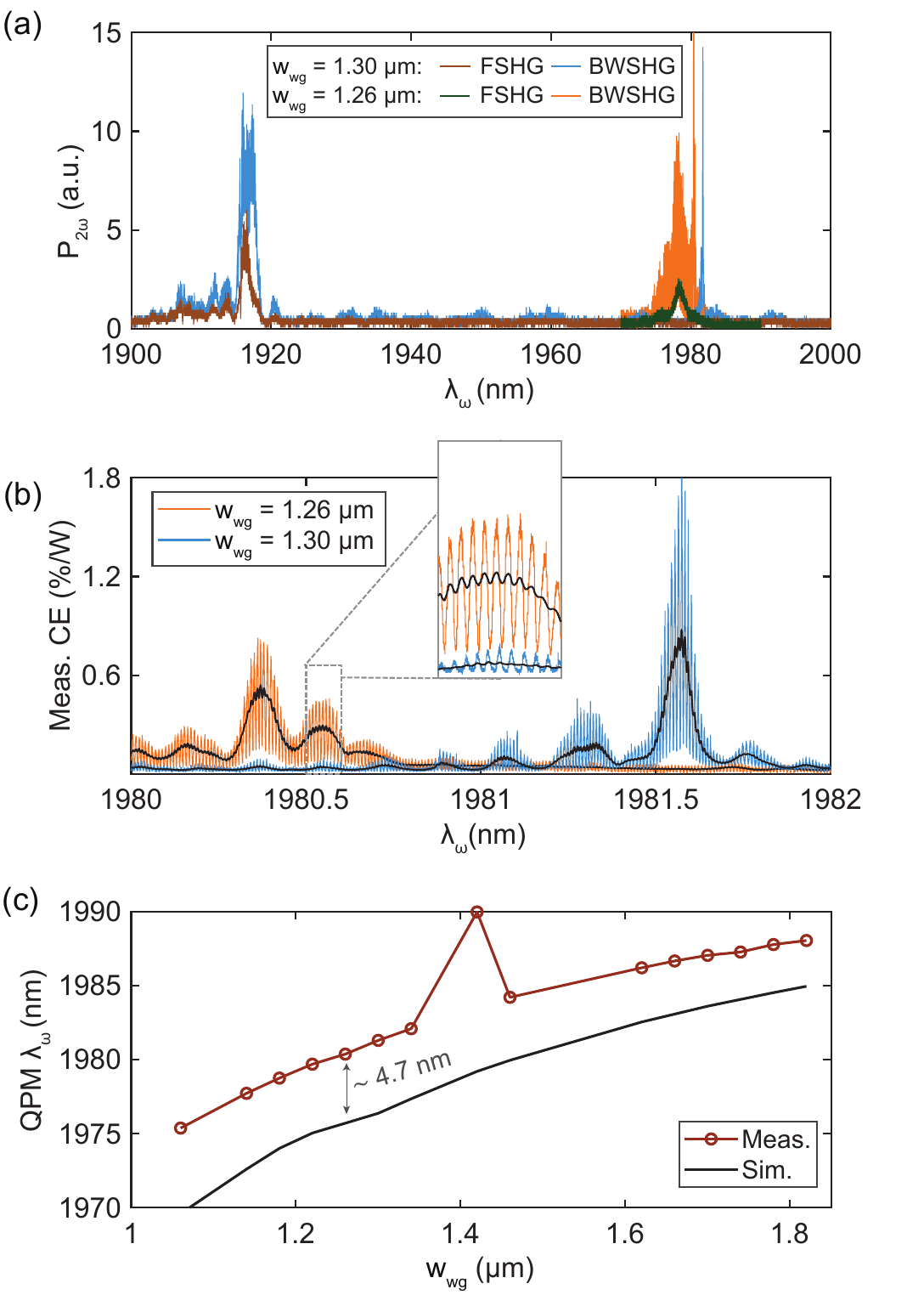}
\caption{\label{fig5} (a) Backward-wave second-harmonic generation (BWSHG) and forward-propagating second-harmonic generation~(FSHG) spectra of the waveguides with design $w_{\rm wg}~=~ $1.30~\textmu m and $w_{\rm wg}~=~ $1.26~\textmu m. (b) Measured BWSHG conversion efficiency (CE) for both waveguides (in orange line~($w_{\rm wg}~=~ $1.26~\textmu m) and blue line ($w_{\rm wg}~=~$1.30~\textmu m)) together with the averaged spectra~(in black). (c)~Measured second harmonic (SH)-peak wavelengths for waveguides with different $w_{\rm wg}$~(brown data points). It is overlaid with the simulated values and indicate a systematic shift of $\sim 4.7$~nm in the peak position.}
\end{figure}

For BWDFG measurements~(Fig.~\ref{fig4}(b)), the same $\sim 1980$~nm light source is coupled from the left arm and 1-ns quasi-CW $\sim 775$~nm source is coupled from the right arm. Dichroic mirrors SP4 and SP5 are placed to monitor the outcoupled input powers through the power meters D4 and D5. In the left arm, SP6 reflects the generated idler toward a lens pair, which minimizes collimation loss. The idler is then monitored with an optical spectrum analyzer (OSA) (Yokogawa AQ6375). For the temperature, pump-wavelength, and incident-power sweeps, a tunable diode laser~(Newfocus TLB-6700) is used instead of the quasi-CW 775~nm source. In the power sweeps, the input power is adjusted through a variable optical attenuator.

\begin{figure*}[!t]
\centering
\includegraphics[width=184.2mm]{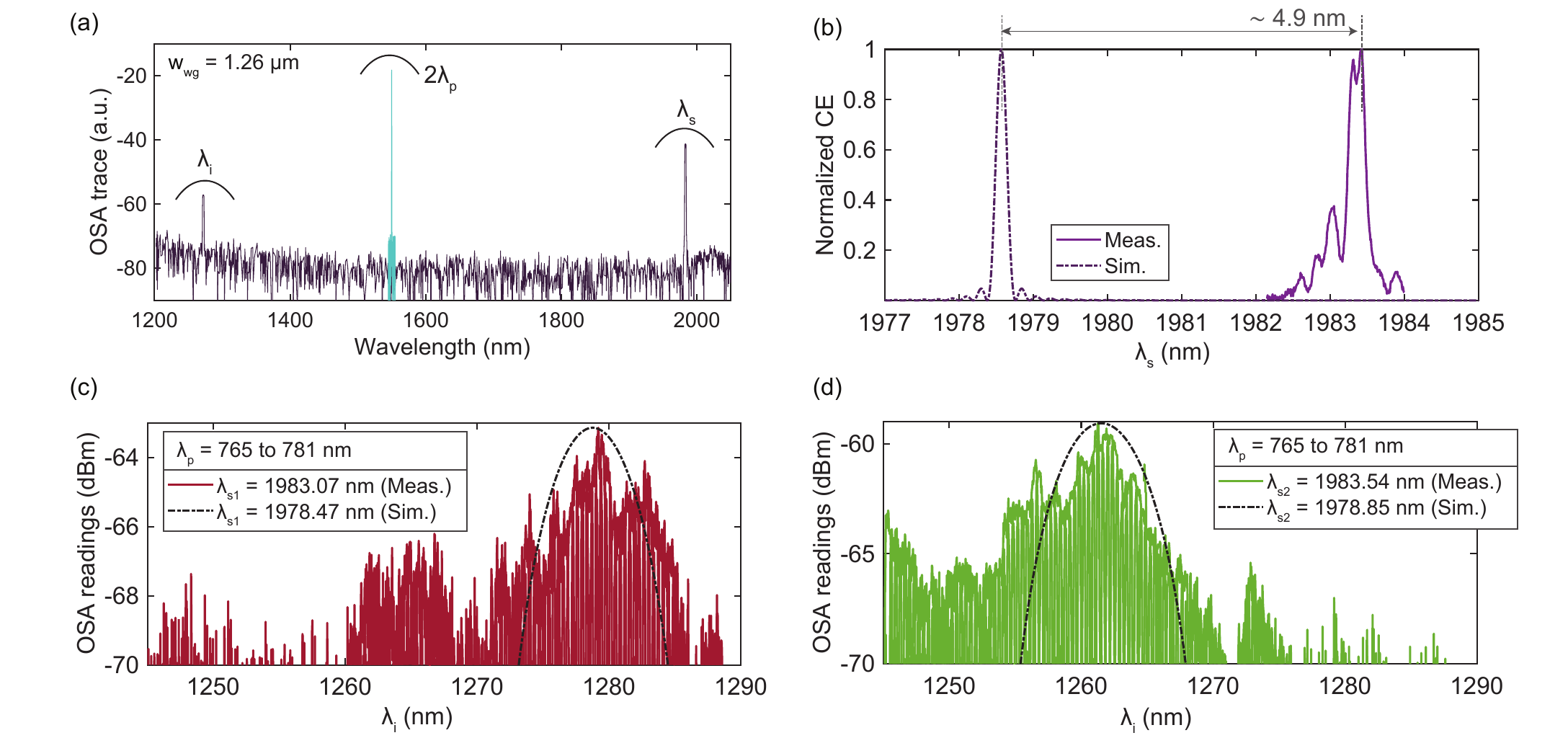}
\caption{\label{fig6} (a)~Optical spectrum analyzer (OSA) spectrum demonstrating signal~($\lambda_{\rm s}$) and idler~($\lambda_{\rm i}$) pair, that is overlaid with the pump~(2$\lambda_{\rm p}$) spectrum. (b)~Solid line:~Normalized measured backward-wave difference-frequency generation~(BWDFG) conversion efficiency (CE) while sweeping $\lambda_{\rm s}$. Dashed line: Normalized simulated CE for BWDFG using nominal design dimensions, namely $w_{\rm wg}$~=~ 1.26~\textmu m, $h_{\rm wg}$~=~800~nm and $\Lambda_{\rm poling}$~=~1425~nm. (c,~d)~Superimposed OSA spectra of the generated idler as a function of measured $\lambda_{\rm i}$ while keeping $\lambda_{\rm s}$ constant at (c)~$\lambda_{s1}$ and at (d)~$\lambda_{s2}$. Black dashed lines: Simulated BWDFG CE, shown in arbitrary units, calculated using \eqref{eq:ce} at the measured peak $\lambda_{\rm i}$ that satisfies both the difference-frequency generation (DFG) quasi-phase matching (QPM) wavelength relation and the QPM condition at $\Lambda_{\rm poling}$~=~1425~nm, in accordance with Fig.~\ref{fig2}a.}
\end{figure*}

\subsection{Experimental Verification of BWSHG} \label{sec_sub:bwshg} 

We present FSHG and BWSHG wavelength scans for two waveguides with different $w_{\rm wg}$ in Fig.~\ref{fig5}(a). In the FSHG scans~(brown line~(for $w_{\rm wg}$~=~1.30~\textmu m) and dark green line~(for $w_{\rm wg}$~=~1.26~\textmu m)), the two waveguides exhibit SH peaks separated by 60~nm, which we attribute to QPM with higher-order SH modes. In BWSHG scans (blue line~(for $w_{\rm wg}$~=~1.30~\textmu m) and orange line~(for $w_{\rm wg}$~=~1.26~\textmu m)), additional spike-like SH peaks appear near 1980~nm with 1~ nm~separation, in good agreement with the designed BWSHG QPM wavelength, demonstrating the insensitivity of this process to $w_{\rm wg}$ offset.

Next, we quantify the measured BWSHG conversion efficiency~(CE) using the pump powers corrected for the dichroic transmission to obtain the outcoupled powers~($P_{\omega_1}$ and $P_{\omega_2}$). CE is then determined using $P_{2\omega}/(P_{\omega_1} P_{\omega_2})$ for both waveguides as shown in Fig.~\ref{fig5}(b) (orange and blue lines), reaching values up to 1.8~\%/W.

Fabry–Perot oscillations arising from the free-space optical path are observed with a periodicity of $\sim 0.01$~nm~(Fig.~\ref{fig5}(b), inset). Averaging is applied in post-processing to extract the spectral envelope with minimal influence from these fluctuations. The filtered spectra of each $w_{\rm wg}$ are shown in black in Fig.~\ref{fig5}(b). The waveguide with $w_{\rm wg}$~=~1.30~\textmu m exhibits a prominent peak with a 3~dB bandwidth $< 0.1$~nm, verifying the narrowband response. Similarly, the waveguide with $w_{\rm wg}$~=~1.26~\textmu m shows a primary SH lobe with a similar bandwidth, accompanied by a pronounced secondary sidelobe, which we attribute to broadening due to poling quality fluctuations.

In Fig.~\ref{fig5}(c), we further extend the measurements to waveguides with different $w_{\rm wg}$ fabricated using the same $\Lambda_{\rm poling}$. The QPM $\lambda_{\omega}$ of each waveguide is marked by brown data points. The outlier corresponds to the waveguide with $w_{\rm wg}$~=~1.42~\textmu m, which exhibits a broadband increasing trend also present in its FSHG spectrum, indicating that the feature originates from forward-propagating SH rather than BWSHG. Excluding this outlier, the remaining data points show a systematic shift of $\sim 5$~nm from the simulated $\lambda_{\omega}$ values phase matched at this $\Lambda_{\rm poling}$~(black line). This shift may originate from the use of literature material parameters, measurement calibration uncertainties, and deviations from the nominal waveguide dimensions.

\subsection{Experimental Verification of BWDFG} \label{sec_sub:bwdfg} 

Next, we characterize the BWDFG response of the waveguide with $w_{\rm wg}$~=~1.26~\textmu m. Fig.~\ref{fig6}(a) shows the OSA spectra of the signal–idler pair at $\lambda_{\rm s}$~=~1983.3~nm and $\lambda_{\rm i}$~=~1271.8 nm, together with the pump spectrum at 2$\lambda_{\rm p}$~=~1549.8~nm. The observed wavelength relation among $\lambda_{\rm p}$, $\lambda_{\rm s}$ and $\lambda_{\rm i}$ confirms the DFG origin of the generated idler.

For further analysis, signal and pump power meter readings are simultaneously monitored on the oscilloscope together with the OSA readings of $\lambda_{\rm i}$. The power meter data are synchronized with the OSA via its trigger output. The outcoupled power immediately after the chip, referred to as $P_{\rm i}$, $P_{\rm p}$ and $P_{\rm s}$, are obtained by scaling the recorded data with the optical path losses. Using this technique, we extract the measured BWDFG CE while sweeping $\lambda_{\rm s}$ (solid line in Fig.~\ref{fig6}(b)).

We highlight that this waveguide's phase-matched BWDFG $\lambda_{\rm s}$~($\sim 1983.4$~nm~(Fig.~\ref{fig6}(b))) is $\sim 3$~nm away from its phase-matched BWSHG $\lambda_{\omega}$~($\sim 1980.4$~nm~(Fig.~\ref{fig5}(b))), experimentally demonstrating
two distinct nonlinear processes within the same waveguide.

Furthermore, in Fig.~\ref{fig6}(b), the simulated BWDFG CE is shown as a dashed line, obtained using the same $w_{\rm wg}$, $h_{\rm wg}$ and $\Lambda_{\rm poling}$ parameter values as Fig.~\ref{fig2}. A similar wavelength offset~($\sim 5$~nm) between the simulated and measured peaks is observed in both the BWDFG and BWSHG experiments, indicating the consistency of the findings across two separate experiments.

Next, we investigate the idler generation while sweeping $\lambda_{\rm p}$ at a constant $\lambda_{\rm s}$. Fig.~\ref{fig6}(c) presents the superimposed OSA spectra measured at $\lambda_{\rm s1}$~=~1983.07 nm for various $\lambda_{\rm p}$ values spanning 765~nm to 781~nm~(top x-axis). The sweep reveals idler generation over the range $\lambda_{\rm i}$~=~1260~nm to $\lambda_{\rm i}$~=~1289~nm. Repeating the sweep at $\lambda_{\rm s2}$~=~1983.54~nm further extends the idler generation toward shorter wavelengths, reaching up to 1244~nm (Fig.~\ref{fig6}(d)). 

To enable a spectral comparison between the measured and simulated responses, we overlay the simulated BWDFG CE in Figs. \ref{fig6}(c) and \ref{fig6}(d). The simulated CE, shown in arbitrary units, is calculated using \eqref{eq:ce} at the measured peak idler wavelength satisfying both the DFG wavelength relation and the QPM condition for 1425 nm poling period. We observe that the measured sidelobes are significantly more pronounced and note that a similar broadening was also present in BWSHG measurements. Nevertheless, a $\sim 0.5$ nm $\lambda_{\rm s}$ tuning produces a measured $\lambda_{\rm i}$ shift of $\sim 17$~nm, consistent with the simulations ($\sim 0.4$~nm $\lambda_{\rm s}$ tuning in the siulations).

\begin{figure}[!t]
\includegraphics[width=92.1mm]{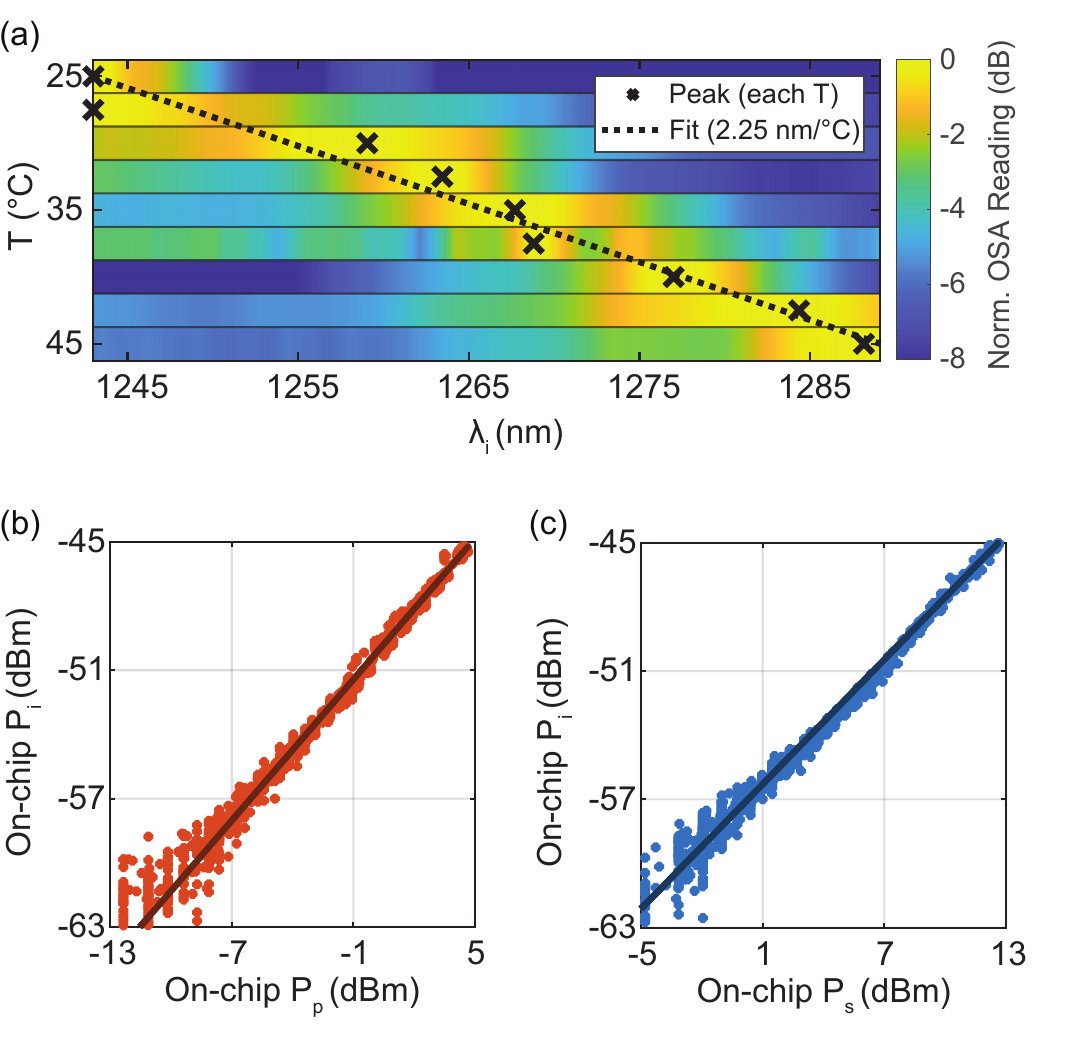} 
\caption{\label{fig7} (a)~Superimposed optical spectrum analyzer (OSA) spectra of the generated idler while sweeping $\lambda_{\rm p}$ at different stage temperatures~(T). (b~and~c)~On-chip power dependence (b)~between $P_{\rm i}$ and $P_{\rm s}$ and (c)~between $P_{\rm i}$ and $P_{\rm p}$ at $\lambda_{\rm s}$~=~1983.28~nm, $\lambda_{\rm p}$~=~775~nm and T~=~40$^\circ$C}%
\end{figure}

As an alternative mechanism for tuning $\lambda_{\rm i}$, we repeat the measurements at stage temperatures ranging from 25$\rm ^{\circ}$C to 45$\rm ^{\circ}$C. After setting each temperature, the input coupling is realigned to minimize variations in coupling loss. To further suppress residual discrepancies, we present normalized superimposed OSA spectra in Fig.~\ref{fig7}(a). By fitting the peak $\lambda_{\rm i}$ values at each temperature (black dashed line), we extract $\Delta\lambda_{\rm i}/\Delta T \rm~=~2.25~\mathrm{nm}/^\circ C$. This enables tuning of $\lambda_{\rm i}$ from 1244~nm to 1290~nm in the investigated waveguides, limited by the $\lambda_{\rm p}$ source rather than the temperature-controlled stage. We emphasize that this tuning range corresponds to a phase-matched $\lambda_{\rm p}$ shift of up to 16~nm while keeping $\lambda_{\rm s}$ fixed.

Finally, we investigate the efficiency of this process, where the measured~CE reaches values up to 0.6~\%/W in the investigated waveguide. We further plot the on-chip $P_{\rm i}$ as a function of the input on-chip powers $P_{\rm p}$~(Fig.~\ref{fig7}(b)) and $P_{\rm s}$~(Fig.~\ref{fig7}(c)) over more than one decade on a log–log scale. The on-chip powers are obtained by correcting for the edge-coupler transmission losses. Linear fits confirm the DFG origin of the process, yielding slopes of 1.1 and 0.97 for the $P_{\rm p}$ and $P_{\rm s}$ sweeps, respectively. We attribute the slight deviation from unity to thermal instabilities at higher input powers, which can increase coupling losses.

\section{Discussion \& Conclusion} \label{sec:conc}

In summary, we demonstrate nondegenerate BWDFG with $\lambda_{\rm s}~\sim~1980$~nm and $\lambda_{\rm i}\sim~1270$~nm, together with BWSHG where $\lambda_{\omega}\sim 1980$~nm. Simulations with nominal design dimensions capture both processes with a QPM wavelength shift of $<~5$~nm, and BWSHG measurements confirm the predicted insensitivity to $w_{\rm wg}$.

Our analysis highlights that tuning through $\lambda_{\omega}$ for BWSHG and $\lambda_{\rm s}$ for BWDFG show comparable behavior, in contrast, $\lambda_{\rm p}$ offers broadband tuning. Experimentally, we achieve tuning from 1244~nm to 1290~nm via $\lambda_{\rm p}$- and temperature tuning. Simulations further project spectral coverage from 1250~nm to 2200~nm through coarse $\lambda_{\rm p}$ tuning~($\sim 300$~nm) complemented by fine $\lambda_{\rm s}$ adjustment~($\leq 6$~nm), determined by the waveguide geometry and further influenced by the group index mismatch between pump and idler.

Measured conversion efficiency remains modest, a limitation primarily attributed to third-order QPM, which imposes a ninefold reduction in peak efficiency, with additional contributions arising from the limited poling depth, duty-cycle errors and missing inverted domains due to the extremely short poling pulses. Nevertheless, these are poling-based limitations rather than fundamental constraints originating from the underlying nonlinear coupled-mode equations \cite{mutterBackwardWaveOptical2024}, and could be mitigated by poling through etched sidewalls \cite{sabattiNanodomainPolingUnlocking2025, frankenMilliwattlevelUVGeneration2026}
or electron-beam-written domains \cite{fergestadCounterpropagatingNondegenerateFrequency2025}, at the cost of increased fabrication complexity.

Our work enables wide-spectrum $\chi^{(2)}$ interactions within a single periodically poled waveguide, opening prospects for large-scale integration. Counter-propagating frequency conversion paves the way for broadband, low-threshold optical parametric oscillation \cite{canaliasMirrorlessOpticalParametric2007, godardBackwardOpticalParametric2022, kasumovaTheoryQuasiphasematchedOptical2024, gaoObservationTuningProperties2025, yangDegeneracylockedOpticalParametric2026} and for on-chip molecular spectroscopy \cite{hwangMidinfraredSpectroscopyBroadly2023, vagberg27MmBackward2024}, while the inherent spatial separation of entangled pairs offers a compelling platform for quantum information applications \cite{luoCounterpropagatingPhotonPair2020, liuObservationFrequencyuncorrelatedPhoton2021, kellnerCounterpropagatingSpontaneousParametric2026, yangSimultaneousType0TypeI2026}.

\begin{backmatter}

\bmsection{Funding} The authors acknowledge Swiss National Science Foundation grants CRSII5-193689 (BLUVES) and BRIDGE-ENABLE (203480) funding. 

\bmsection{Acknowledgement} Fabrication of the lithium niobate waveguides was carried out at the Center of MicroNanoTechnology (CMi) at EFPL. Two-photon imaging was performed in the UNIL Cellular Imaging Facility (CIF).

\bmsection{Disclosures} The authors declare no conflicts of interest.

\end{backmatter}

\bibliography{BetterBib}

\end{document}